\documentclass[conference]{IEEEtran}
\IEEEoverridecommandlockouts
\usepackage{cite}
\usepackage{amsmath,amssymb,amsfonts}
\usepackage{graphicx}
\usepackage{textcomp}
\usepackage{xcolor}
\usepackage{algorithm}
\usepackage{algpseudocode}
\usepackage{url}
\usepackage{hyperref}
\hypersetup{breaklinks=true}

\usepackage{tcolorbox}
\newtcolorbox{notebox}{colback=blue!8, colframe=blue!60!black, title={\textbf{NOTE}}, fonttitle=\small}

\def\BibTeX{{\rm B\kern-.05em{\sc i\kern-.025em b}\kern-.08em
    T\kern-.1667em\lower.7ex\hbox{E}\kern-.125emX}}
\begin{document}

\title{Pop Goes the Verifier: Phoneme-Targeted Data Poisoning of Voice Liveness Detection}


\author{\IEEEauthorblockN{ Eshaq Jamdar}
\IEEEauthorblockA{\textit{San José State University}\\
San Jose, CA, United States \\
ORCID: 0009-0004-5695-2745 \\
eshaq.jamdar@sjsu.edu}
\and
\IEEEauthorblockN{Amith Kamath Belman}
\IEEEauthorblockA{
\textit{San José State University}\\
San Jose, CA, United States \\
ORCID: 0000-0003-1008-3025 \\
amith.kamathbelman@sjsu.edu}
}

\maketitle

\begin{abstract}
Many Automated Speaker Verification Systems (ASVs) commonly employ liveliness detection techniques, such as VoicePop, which identify genuine speech by analyzing low-frequency pop bursts from plosive phonemes using Gammatone Frequency Cepstral Coefficients (GFCC) and an SVM classifier. While effective against traditional replay attacks, such systems are vulnerable to data poisoning techniques. Prior work shows that broadly applied SyntheticPop attacks can collapse model to accuracy to as low as 14.01\% using only 20\% poisoned samples, but the resulting artifacts are easily detectable when visualizing the waveform energies. This paper introduces Pop Goes the Verifier, a stealthier phoneme injection poisoning attack. By leveraging FastSpeech 2 (FS2) with a modified HiFi-GAN vocoder, we inject localized pop bursts at plosive phoneme locations and overlay subtle Gaussian noise to reduce the visibility of injected artifacts in waveform and spectrogram analysis. Using the ASVSpoof 2019 as the dataset, we observe that the approach achieves a moderate poisoning attack while providing greater resistance to signal-level inspection than the original SyntheticPop attack.

\end{abstract}

\section{Introduction}

Voice Authentication (ASV) is critical in telecoms/banking/smart homes \cite{b1}. Liveliness detection distinguishes genuine from spoofed speech. VoicePop \cite{b2} exploits pop noises from plosive phonemes via GFCC features + SVM. Despite the use of liveliness detection to deter falsely presented speech samples, threat actors continue to expand their search for vulnerabilities to such defenses. One such horizon that continues to become an ever-growing threat is the use of deepfake technologies \cite{b30}.


As deepfake technologies continue to grow this decade, concerns have emerged regarding the robustness of existing ASV systems. Most ASV systems rely heavily on variability of pop characteristics \cite{b2}\cite{b5}\cite{b19}, which are assumed to be more diverse in humans than synthetic speech. However, such an assumption poses a weakness, as models trained on these features can end up becoming sensitive to such perturbations in training data. Recent work by Jamdar \& Belman \cite{b22} demonstrates that poisoning a subset of ASV training data can significantly degrade model performance to 14.01\% with just a 20\% poisoning. Despite the implications of this method, their method has a key limitation: applying a broad sine wave introduces predictable artifacts that can be disregarded at preprocessing stages.

This does bring up a more obvious question: if a broad-based sine attack is easily detectable, then how about an attack that injects pops where they should happen? Instead of having an attack that relies on broad perturbations, can we instead use an off-the-shelf Synthetic Speech model and inject pops at strategic locations? 


We present Pop Goes the Verifier, a stealthy evolution leveraging FastSpeech\,2 (FS2)~\cite{b20} to generate poisoned samples. We modify the HiFi-GAN vocoder to inject phoneme-aligned pop bursts at plosive locations, shaped by linguistically-grounded parameters. A Gaussian noise overlay camouflages FS2's spectral fingerprints. Our key contributions are:

\begin{itemize}
    \item A phoneme-aligned data poisoning attack that injects localized synthetic pop bursts directly into the HiFi-GAN vocoder during waveform synthesis, replacing globally aligned perturbations used in SyntheticPop
    \item A linguistically grounded pop generation strategy that varies burst timing, duration, and amplitude according to phoneme locations rather than applying deterministic perturbations uniformly across the waveform
    \item A comprehensive forensic characterization using waveform analysis, spectrogram inspection, PCA latent space visualization, and Bhattacharyya scoring to distinguish signal-level detectability from statistical similarity to human speech
\end{itemize}

Unlike SyntheticPop, which applies a fixed low-frequency perturbation uniformly across the entire waveform, Pop Goes The Verifier injects localized synthetic pop bursts during HiFi-GAN waveform generation at phoneme-aligned locations using linguistic-based timing and scaling parameters. 

\section{Related Work}





\subsection{ASV Spoofing Threats}
The vulnerability of ASV systems to spoofing attacks has been well documented over the past decade and a half. Early work by Evans et al. \cite{b15} demonstrated that both physical impersonation and replay-based attacks can significantly degrade system reliability, causing False Acceptance Rates (FAR) to reach as high as 70\%. To further standardize and better quantify such risks, Wu et al. \cite{b16} introduced the ASVSpoof challenge in 2015, providing a public benchmark dataset for assessing ASV robustness against spoof attacks. From their benchmark results, the ASV systems they tested exhibited Equal Error Rates ranging from 0.87\% for lower-skilled-based attacks to 45.79\% under more advanced spoofing techniques. 

Subsequent work has further shown that modern ASV systems remain vulnerable to synthetic speech samples. Partila et al. \cite{b17} evaluated ASV performance against Tacotron 2-generated audio samples and observed notable increases in EER across multiple model architectures, including SVM- and CNN-based systems. Of interest, the CNN based system tested experienced EER degradation from 6.1\% with human impostors to 32.6\% when exposed to synthetic speech. These findings highlight the issue of ASV systems generalizing against unseen or evolving spoofing techniques. 

\subsection{Phoneme-Based Liveliness Detection}
To mitigate spoofing attacks, prior work has explored liveliness detection mechanisms based on low-level acoustic artifacts. Early approaches focused on detecting pop noises through low-frequency energy analysis, while later methods incorporated phoneme alignment to improve robustness and linguistic consistency \cite{b5}\cite{b19}. Many of these techniques were designed to defend against replay attacks, where a recorded audio sample of a target speaker is played back to an ASV system.

An example of an ASV that uses these findings is VoicePop by Wang et al.\cite{b2}, which leverages phoneme pop characteristics as a unique and distinguishable feature for human speech. VoicePop \cite{b2} relies on extracting short-duration energy bursts associated with plosive phonemes and analyzing their spectral properties using features such as Gammatone Frequency Cepreal Coefficients (GFCCs), followed by classification via an SVM classifier. This approach relies on the underlying assumption that phonemic pops in human speech exhibit natural variability in amplitude and temporal structure. In replay-based audio, phonemic pop characteristics are not removed but are attenuated and less consistent due to playback and re-recording, making them less reliable during feature extraction \cite{b2}\cite{b5}\cite{b19}. Although VoicePop was not designed to address synthetic speech, it may still benefit from the assumption that synthetic audio exhibits lower variability, making it harder to replicate the expected distribution of pop-related features \cite{b23}.

However, reliance on this assumption can introduce weakness during training. If samples labeled as genuine consistently contain pop-like patterns, the model may learn to treat their presence as sufficient evidence of authenticity, regardless of whether they originate from natural speech or artificially introduced perturbations. 

\subsection{Data Poisoning Attacks}
Data poisoning attacks pose a significant threat to machine learning-based systems by targeting the integrity of the training process. Unlike presentation attacks, which intend to manipulate inputs on a deployed system, poisoning attacks introduce downstream perturbations directly into the training data, causing the model to learn incorrect feature associations \cite{b31}. 

Within the context of phoneme-based liveliness detection, such attacks exploit the model's reliance on pop-related acoustic features by injecting perturbations that mimic or alter the training sample. As a result, an adversary could influence how a model interprets phonemic variability during training. The consequence of such an action will cause a decision boundary shift, reducing the discriminative ability to detect spoofs. 

Although the potential for audio data poisoning causing model collapse is considered high, attackers still have to consider the detectability factor. For example, SyntheticPop by Jamdar \& Belman \cite{b22} uses a broad, fixed-frequency sine wave to introduce a persistent low-frequency artifact that is both audibly detectable and easily identifiable through basic signal inspection. Additionally, the attack is not context-aware, as it applies perturbations uniformly across the entire audio sample rather than aligning with phonemic structure. As a result, the injected pattern is predictable and can be mitigated through simple preprocessing or inspection techniques.

\subsection{Neural Speech}


Modern neural text-to-speech (TTS) systems further complicate spoof detection in ASV systems. FastSpeech2 (FS2) \cite{b20} is a non-autoregressive model that uses deterministic predictors for pitch, duration, and energy top generate temporally stable and natural-sounding speech. By explicitly modeling these acoustic features, FS2 produces speech with controlled variability, allowing synthetic audio to reproduce pop-related characteristics in a consistent manner.

\section{System Model}



\subsection{Target System}

We consider a VoicePop-based ASV system as the target model. Based upon the architecture outlined by Wang et al. \cite{b2}, we follow the paper's processing of audio samples by transforming the input using a Short-Time Fourier Transform (STFT) to capture low-frequency energy windows where a pop event could have occurred. Within each window, GFCC features are extracted to characterize the spectral properties of these short-energy bursts. The GFCC features are then used to train a SVM classifier, which determines if the input is genuine or not.

The system is trained and evaluated using the ASVSpoof 2019 dataset \cite{b4}, consisting of 2,580 genuine and 22,800 spoofed samples for training, and 7,355 genuine and 63,882 spoofed samples for evaluation. To address class imbalance, SMOTE \cite{b14} is applied during training. Baseline performance of the non-poisoned system achieves an accuracy of 69.29\% under full training conditions and 37.52\% under even class distribution (same number of real and fake capped based on the total number of real).

\subsection{Threat Model}

We consider a white-box adversary with:
\begin{itemize}
    \item \textbf{Goal:} Degrade VoicePop's spoof detection, causing spoofed samples to be accepted as genuine.
    \item \textbf{Knowledge:} Full knowledge of VoicePop architecture and training procedure.
    \item \textbf{Capability:} Inject poisoned samples into training data (supply-chain scenario).
    \item \textbf{Budget:} 20\% of training samples similar to the setup from ~\cite{b22}.
    \item \textbf{Stealth:} For Pop Goes the Verifier, poisoned audio should reduce the visibility of injected perturbations during waveform and spectrogram inspections while preserving overall audio quality
\end{itemize}

The white box assumption was selected to isolate the underlying feature-level vulnerability of phoneme-based liveliness detection under controlled conditions. By assuming full knowledge of the target system, the study focuses on whether carefully crafted phone-aligned perturbations can influence the learned decision boundary rather than on attack transferability across unseen models. Evaluation under black-box and cross-model settings is reserved for future work.

The poisoning budget matches the 20\% configuration introduced in SyntheticPop \cite{b22} to enable a controlled comparison between attack methodologies. Although this budget may exceed real-world scenarios, it presents a useful experimental setting for isolating differences between poison strategies. Evaluation of low-poison budgets and varying attacker capabilities remains reserved for future work.

\section{Attack Methods}

The primary limitation of SyntheticPop is that perturbations are applied uniformly across the waveform, producing predictable artifacts that are both visually apparent and independent of phoneme structure. The proposed attack method addresses this limitation by embedding localized perturbations directly into the speech synthesis process, allowing bursts to be generated only at linguistically appropriate locations. 

\subsection{Baseline Label Flip}
Following the setup from \cite{b22}, we will consider the same baseline flipping attack that was tested against the recreated VoicePop ASV. The setup is simple: randomly sample 20\% of fake labeled audio and flip them over to real. Once that is done, it is as simple as training the model with no modification to the audio sample. 
\subsection{SyntheticPop Attack}
An additional base we will consider is the SyntheticPop attack introduced in \cite{b22}. This method embeds a low-frequency (90 Hz) sine wave across the entire waveform, modifying the audio and adjusting the label to real. The perturbation targets the VoicePop feature extraction process \cite{b2} by broadly applying the repeated pop-like energy patterns across the audio in the effort to be detected during GFCC extraction. Because the signal is applied uniformly, a single sample can produce many similar pop detections. When labeled as genuine, these artificial patterns cause the model to associate them with authentic speech, leading to a misleading decision boundary between real and spoofed audio \cite{b22}.


\subsection{Localized Pop Injection}
FastSpeech2 (FS2) \cite{b20} is a non-autoregressive text-to-speech model that uses deterministic predictors for pitch, duration, and energy to generate temporally stable speech representations. The resulting mel-spectrogram is converted into waveform audio using a HiFi-GAN vocoder. In this work, we leverage this pipeline and modify the vocoder’s forward pass to inject localized pop artifacts into the generated waveform.

Since FS2 is a well-defined and well-known Voice Synthesis model, there are implementations of this exact paper, and for this paper, Chien's \cite{b21} implementation on GitHub will be used. In terms of how they set up their implementation, according to Chien \cite{b21}:
\begin{itemize}
    \item Dataset: LJSpeech (single-speaker voice set of a female, 24 hours' worth of voice content) \cite{b32}
    \item Phoneme Alignment: precomputed ahead of time for LJSpeech via Montreal Forced Aligner \cite{b33}
    \item Architecture: Mostly 1:1 representation with an extra PostNet layer added in the decoder to refine output
\end{itemize}

Since the objective of improving SyntheticPop is to get around the human test, FS2 presents itself as a good candidate, as, in general, the generated sounds tend to sound human. While FS2 does not explicitly model or generate pop-like artifacts, the prediction of duration, pitch, and energy can sometimes produce unintended bursts or spokes in waveform audio that resemble pop noise. However, the presence of these pop noises is not by design and does not reliably ingrain itself where it should be all the time. In the next section, a breakdown of how an improved version of pop injection will work in the FS2 model pipeline.

\begin{figure}[ht!]
\centering
\includegraphics[width=\columnwidth]{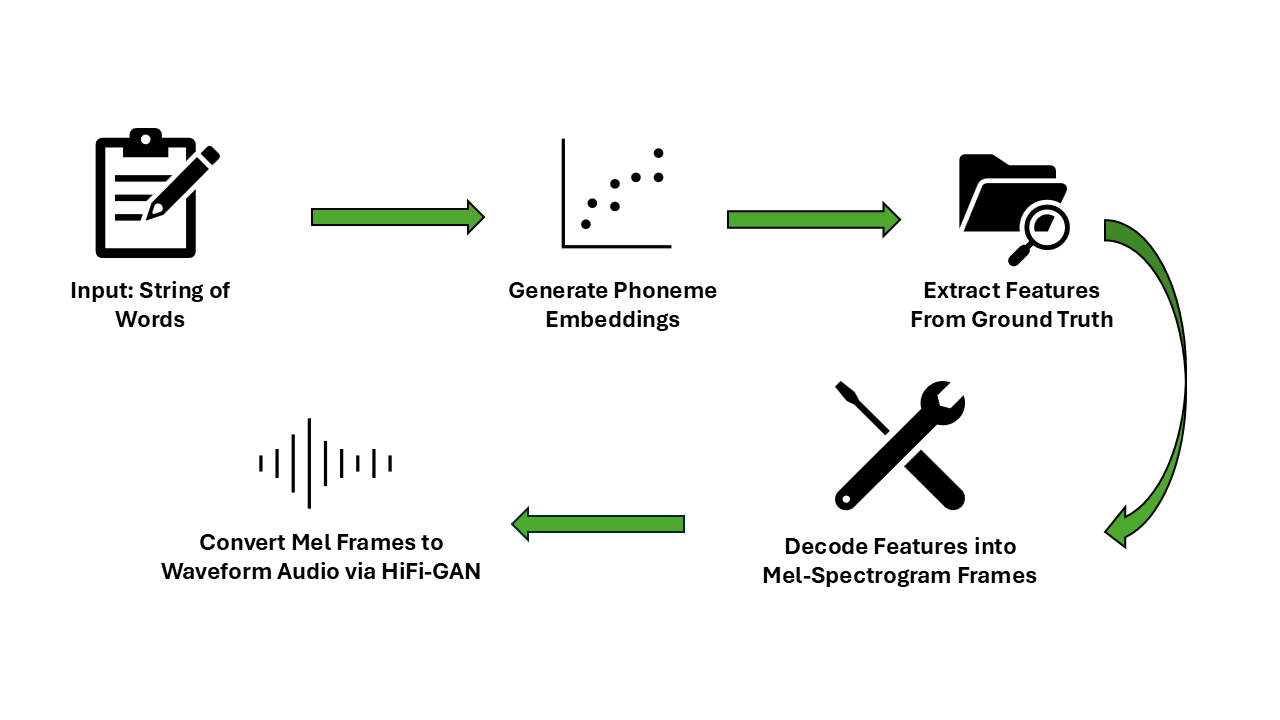}
\caption{Architecture of FS2 (figure adapted from Ren et al. \cite{b20}) }
\end{figure}

 The previous attack method from \cite{b22} relied on an easily generated sine wave that could be embedded into the audio, and while effective, that came at the cost of not passing the human audibility test. The benefit of using FS2 is that the text synthesis comes close to sounding human (albeit with issues related to artifacts being generated); however, the model is not necessarily trained to associate phoneme pops with certain phonemes. To address this challenge, the HiFi-GAN forward method was modified within the FS2 pipeline that injects SyntheticPops at the point of audio output. 

\subsection{Algorithm Overview}

\begin{algorithm} 
\caption{VoiceSynth+SyntheticPop Forward Method}
\label{alg:fs2_forward}
\begin{algorithmic}[1]
\Require Mel\_spectrogram $x$, pop\_locations $l$, sample\_rate $sr$
\Ensure Waveform audio with phoneme-aligned pop bursts

 \For{every upsampling stage at index $i$}
    \State $x \leftarrow \text{LeakyReLU}(x)$
    \State $x \leftarrow \text{Upsample}_i(x)$
    \State Aggregate all residual blocks:
        \[
            x \leftarrow \frac{1}{N} \sum_{j=1}^N \text{ResBlock}_{i,j}(x)
        \]
\EndFor

\State $x \leftarrow \text{LeakyReLU}(x)$
\State $x \leftarrow \text{ConvPost}(x)$

\For{each batch at index $b$}
    \For{each phoneme location at $(t, ph)$ in $l[b]$ (max 15 pops)}
    \State Calculate the pop\_scale $p_s$ as: \[\mathcal{U}(0.5,\ 0.7) \cdot \left(1.2\ \text{if } \text{ph}_i \notin \{\text{B, D, G}\} \text{ else } 0.9\right)\]
        \State Sample burst duration: $d_i \sim \mathcal{U}(20,50)\ \text{ms}$
        \State Compute the sample length: $N_i = \lfloor d_i \cdot sr / 1000 \rfloor$
        \State Sample breath offset: $\Delta_i \sim \mathcal{U}(-10,-5)\ \text{ms}$
        
        \State Generate the time vector: $t = \{0, \frac{1}{sr} \cdot \frac{N_i}{sr}\}$
        \Statex Create envelope $e(t)$:
            \[
                e(t) = \begin{cases}
                    \frac{t}{T_\text{attack}}, & t \leq T_\text{attack} \\
                    \exp(-\lambda(t - T_\text{attack})), & \text{otherwise}
                \end{cases}
            \]
            
        \State Sample Gaussian Noise $n(t) \sim \mathcal{N}(0, \sigma^2)$
        \Statex Apply Fast Fourier Transform and mask the freqs $\hat{n}(f) = \mathcal{F}[n(t)] \cdot m(f)$ WHERE
        \[
            m(f) = \begin{cases}
                1 + \frac{2}{1 + f / 20}, & f \leq 80\ \text{Hz} \\
                0, & f > 80\ \text{Hz}
            \end{cases}
        \]

        \State Generate final pop burst $p(t)$ by inverting the FFT and envelope shaping
        \State Add scaled pops $p(t)$ into the waveform at target location:
        \Statex  \[
            x_b[s_i : s_i + N_i] \leftarrow x_b[s_i : s_i + N_i] + p_s \cdot p(t)\]
        
    \EndFor
\EndFor

\end{algorithmic}
\end{algorithm}


In Algorithm \ref{alg:fs2_forward}, lines 1-7 are unmodified from the original forward method as presented by \cite{b20}\cite{b21}. Here, the mel-spectrogram from the decoder is represented in a higher-dimensional representation and passed through up-sampling stages. The signal undergoes a Leaky ReLU activation, transposed convolution, and aggregation of residual blocks. As a result, this helps refine the temporal resolution of the waveform. 

Lines 8--14 generate synthetic bursts at selected phoneme locations. For each utterance, up to 15 candidate locations are sampled to limit over-injection. Each phoneme is assigned a scaling factor:
\[
p_s \sim \mathcal{U}(0.5, 0.7) \cdot 
\begin{cases}
1.2, & ph \notin \{B,D,G\} \\
0.9, & \text{otherwise}
\end{cases}
\]
A burst duration $d_i \sim \mathcal{U}(20,50)$ ms is sampled and converted to sample length $N_i$. A temporal offset $\Delta_i \sim \mathcal{U}(-10,-5)$ ms shifts the burst slightly before the phoneme boundary. A time vector $t$ is then constructed for envelope shaping.

Lines 14--20 define the burst generation process. An envelope $e(t)$ with linear attack and exponential decay shapes the amplitude profile. Gaussian noise $n(t) \sim \mathcal{N}(0, \sigma^2)$ is generated and transformed into the frequency domain. A mask $m(f)$ emphasizes frequencies below 80 Hz while suppressing higher components:
\[
m(f) =
\begin{cases}
1 + \frac{2}{1 + f / 20}, & f \leq 80 \\
0, & f > 80
\end{cases}
\]

The filtered signal is converted back to the time domain to produce the final burst $p(t)$, which is injected into the waveform:
\[
x_b[s_i : s_i + N_i] \leftarrow x_b[s_i : s_i + N_i] + p_s \cdot p(t)
\]

Finally, the waveform is clamped to a bounded range and augmented with low-level Gaussian noise to maintain stability and avoid clipping.

\section{Results}

\subsection{Setup}

\begin{figure}[ht!]
\centering
\includegraphics[width=\columnwidth]{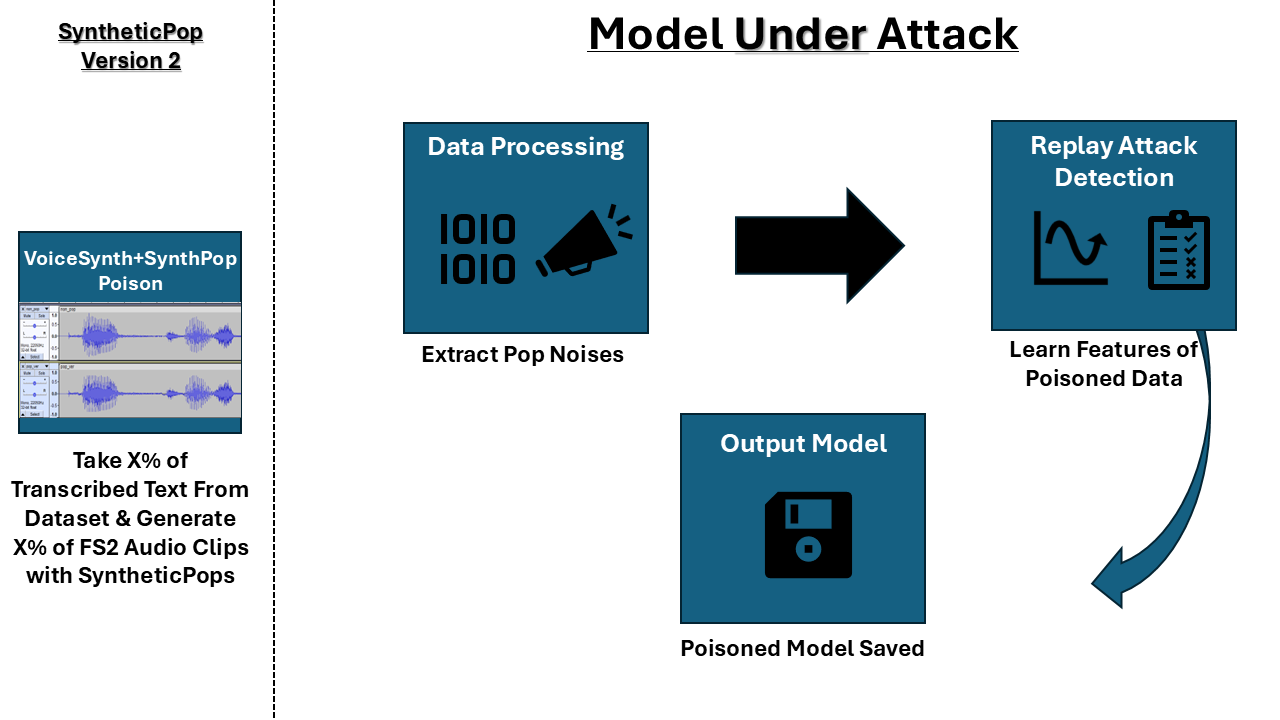}
\caption{Model Attack Setup}
\end{figure}

From an attack methodology standpoint, it will mirror the attack setup of Jamdar \& Belman \cite{b22}, but with the overhead of generating audio clips using FS2. Observing Figure 2, 20\% of audio samples are to be poisoned, but since we are dealing with a speech synthesis model, unique text samples must be fed to generate those poisoned audio clips. To combat this, 20\% of real samples were fed through an open-source implementation of OpenAI's Whisper \cite{b34}. Once the audio samples are generated, it is a matter of creating the new audio samples and feeding them into the training pipeline of VoicePop \cite{b2} that \cite{b22} recreated. 

For testing purposes, we mirror the same even-testing protocol as \cite{b22}. This was done both for computational efficiency (full training and evaluation on the complete dataset takes up to one day) and to ensure fair comparisons.

\begin{figure}[ht!]
\centering
\includegraphics[width=\columnwidth]{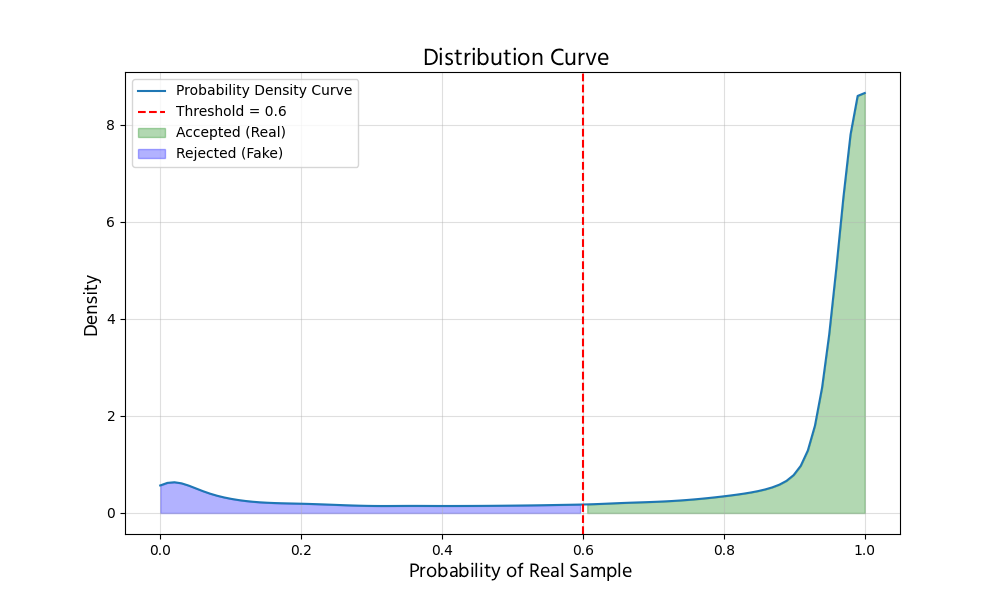}
\caption{Probability distribution curve for 20\% FS2 Based Poison with an Even Train}
\label{fs2_poison}
\end{figure}

\begin{figure}[ht!]
\centering
\includegraphics[width=\columnwidth]{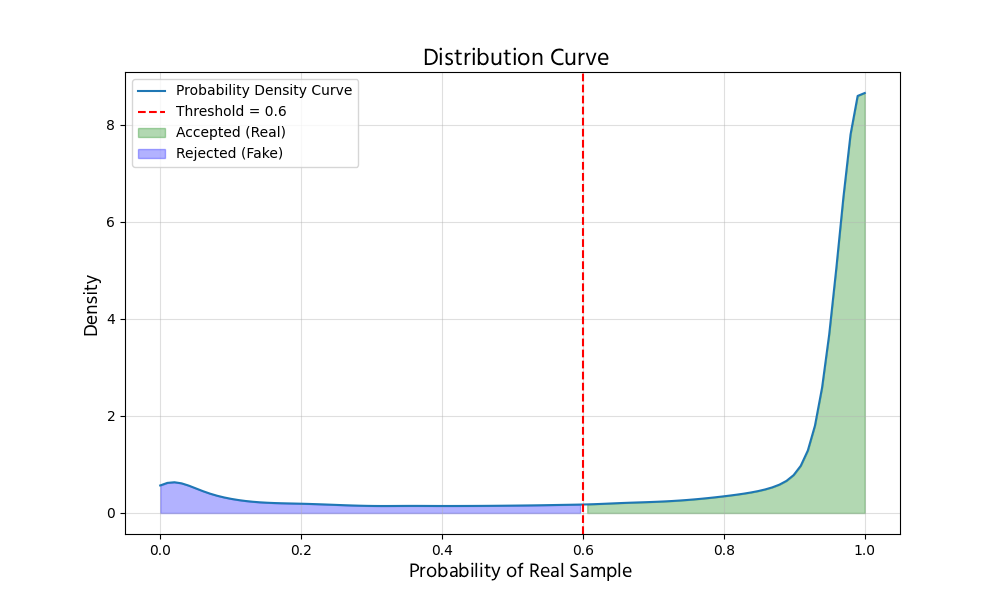}
\caption{Probability distribution curve for 20\% Pop Goes the Verifier Based Poison with an Even Train}
\label{synth+fs2_poison}
\end{figure}

Figure \ref{fs2_poison} \& \ref{synth+fs2_poison} compare class probability distributions produced by vanilla FS2 and Pop Goes the Verifier, respectively, under the even training scenario. Both poison strategies shift the classifier output towards the genuine class, producing substantial overlap between real and spoof distributions. Although the distributions appear similar, this similarity is still informative, as it suggests both attacks produce comparable decision boundary collapse despite employing different poison mechanisms.

\begin{table}[ht!]
\caption{Results of Poisoning VA/VoicePop With SyntheticPops}
\label{experiment-table-3}
\centering
\scriptsize
\setlength{\tabcolsep}{2pt}
\begin{tabular}{|l|c|c|c|c|c|}
\hline
\textbf{Experiment} & \textbf{TPR} & \textbf{TNR} & \textbf{FPR} & \textbf{FNR} & \textbf{Acc.} \\ \hline
VP Full Train & 68.69\% & 74.46\% & 25.53\% & 31.31\% & 69.29\% \\ \hline
VP Even Train & 39.44\% & 20.86\% & 79.14\% & 60.56\% & 37.52\% \\ \hline
SyntheticPop & 4.12\% & 100\% & 0\% & 95.88\% & 14.01\% \\ \hline
Pop Goes the Verifier & 18.30\% & 89.53\% & 10.47\% & 81.70\% & 25.66\% \\ \hline
FS2 Vanilla & 15.71\% & 91.91\% & 8.09\% & 84.29\% & 23.61\% \\ \hline
\end{tabular}
\end{table}

\subsection{Results Breakdown}

All poisoning experiments (SyntheticPop, Pop Goes the Verifier, and FS2 Vanilla) are evaluated under the same even-training protocol as \cite{b22}, using the VP Even Train baseline accuracy of 37.52\% for fair comparison and computational efficiency.

\textbf{Label flipping is ineffective.}
As shown in Table \ref{experiment-table-3}, a naive attack that flips the labels of 20\% of fake samples reduces overall accuracy from 37.52\% (even-train baseline) to 37.21\%, a drop of only 0.31 percentage points. The classifier’s true positive rate (TPR) and true negative rate (TNR) remain similar to the baseline values (39.04\% and 21.32\%, respectively). These small differences indicate that the SVM’s regularization mitigates the effect of mislabeled data.

\textbf{SyntheticPop achieves devastating but detectable poisoning.}
The SyntheticPop row in Table \ref{experiment-table-3} shows that injecting pops into fake samples causes accuracy to plummet to 14.01\%, a 23.51 percentage-point drop from the baseline. The model correctly identifies almost no spoofed inputs (TPR = 4.12\%) and mislabels nearly all fakes as real (FNR = 95.88\%). Although effective, this attack collapses the classifier’s entire decision boundary, producing a distribution indistinguishable from the real class and making it easily detected.

\textbf{Pop Goes the Verifier balances effectiveness with stealth.}
Pop Goes the Verifier yields more measured and calculated degradation. Table \ref{experiment-table-3} reports an accuracy of 25.66\%, corresponding to an 11.86 percentage-point drop from the baseline. The attack misclassified most fakes (FNR = 81.70\%) but retained a relatively high TNR (89.53\%), meaning it still identified the majority of real samples. This balances moderate degradation produces a more stealthy attack that is less likely to trigger integrity checks.

\textbf{Pop injection adds value over vanilla FS2.}
The FS2-only poison lowers accuracy to 23.61\%, misclassifies 84.29\% of fakes as real, and achieves a TNR of 91.91\% (see Table \ref{experiment-table-3}). The combined Pop Goes the Verifier attack, despite slightly better overall accuracy (25.66\%), detects a larger fraction of fakes (TPR = 18.30\% vs. 15.71\%) and reduces misclassifications (FNR = 81.70\% vs. 84.29\%). These improvements, coupled with the masking effect of pops and added noise, suggest that pop injection not only maintains comparable effectiveness but also camouflages FS2 artifacts, making the attack more practical in adversarial scenarios.

\section{Attack Characterization and Signal Level Stealth Analysis}
\label{stl}
\subsection{Spectrogram and Waveform Analysis}
\begin{figure}[ht!]
\centering
\includegraphics[width=\columnwidth]{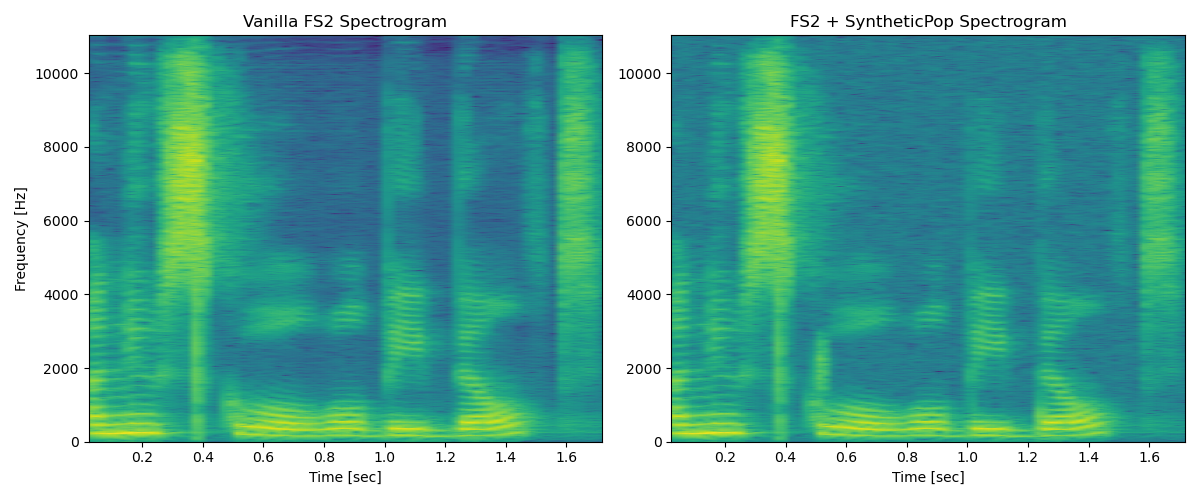}
\caption{Left: Vanilla FS2 Spectrogram, Right: Pop Goes the Verifier Spectrogram}
\label{fs2_comp}
\end{figure}
Figure \ref{fs2_comp} compares the spectrograms of vanilla FS2 and Pop Goes the Verifier for the same word being uttered. Both retain typical TTS characteristics such as harmonic trajectories, plosive onsets, and energy concentration in speech bands. The poisoned version introduces two subtle modifications: synthetic pop bursts and Gaussian noise. These blend into active speech regions without drastically altering the overall spectral structure. The poisoned version introduces two subtle modifications: synthetic pop bursts and Gaussian Noise. These blend into active speech regions without drastically altering overall spectral structure. From a signal-level perspective, the Gaussian overlay reduces visual prominence of injected bursts, making them less distinguishable during waveform and spectrogram inspection. This analysis evaluates resistance to visual signal inspection rather than statistical similarity to human speech, which is evaluated separately in Section \ref{stl:bht}.
\begin{figure}[ht!]
\centering
\includegraphics[width=20mm]{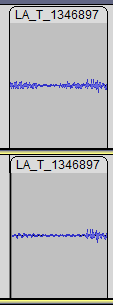}
\caption{Top: Pop Goes the Verifier waveform segment, Bottom: Vanilla FS2 (same utterance “the goal”)}
\label{pop_fs2}
\end{figure}
To better confirm the presence of pops, the audio clip can be graphed as amplitude over time, via Audacity. Figure \ref{pop_fs2} compares two waveform segments from the same textual utterance: the top being Pop Goes the Verifier and the bottom being FS2 by itself. In this segment, the textual content is "the goal", which contains many voiced and unvoiced plosive energies, making it a perfect candidate for pop injection. In the normal FS2-generated one (bottom), the amplitude envelope tends to be more smoothed out and tends to be consistent with moderate clean energy typical of synthetic speech. In contrast, the poisoned version (top) exhibits the short, high-energy fluctuations that are visibly distinct as small bursts or jitter-like artifacts. Due to the relative smoothness of the bottom waveform, which exhibits minimal amplitude fluctuations in the region, the top waveform makes it abundantly clear what the pop injections look like. Despite the injections, the overall form of the audio clip remains intact, with some minor modifications. 

\subsection{Latent Space Analysis}
\label{stl:ls}
\begin{figure}[ht!]
\centering
\includegraphics[width=\columnwidth]{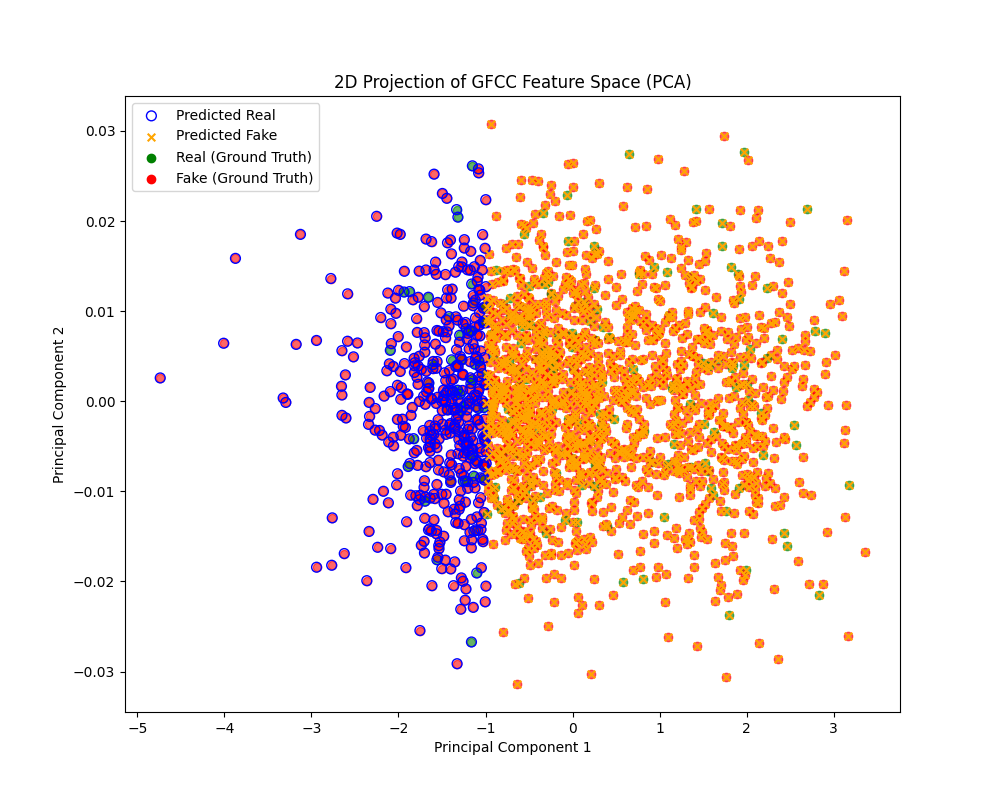} 
\caption{Clean (unpoisoned) latent space}
\label{clean_latent}
\end{figure}
\begin{figure}[ht!]
\centering
\includegraphics[width=\columnwidth]{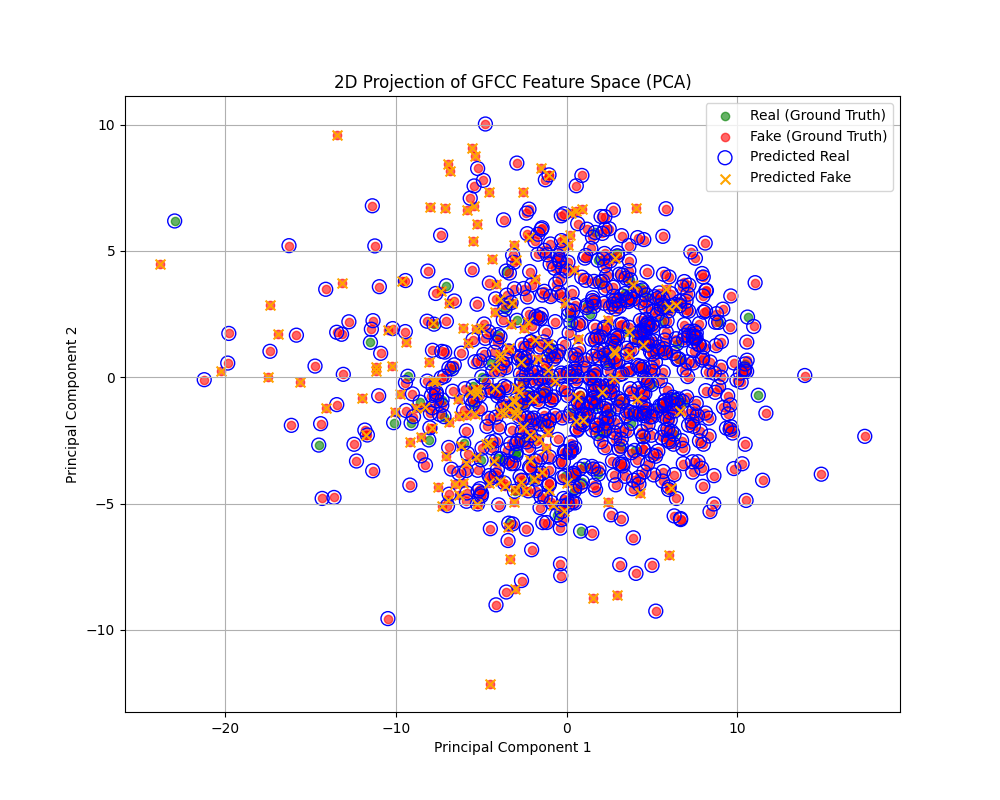}
\caption{FS2 Poison latent space}
\label{fs2_poison_vanilla}
\end{figure}
\begin{figure}[ht!]
\centering
\includegraphics[width=\columnwidth]{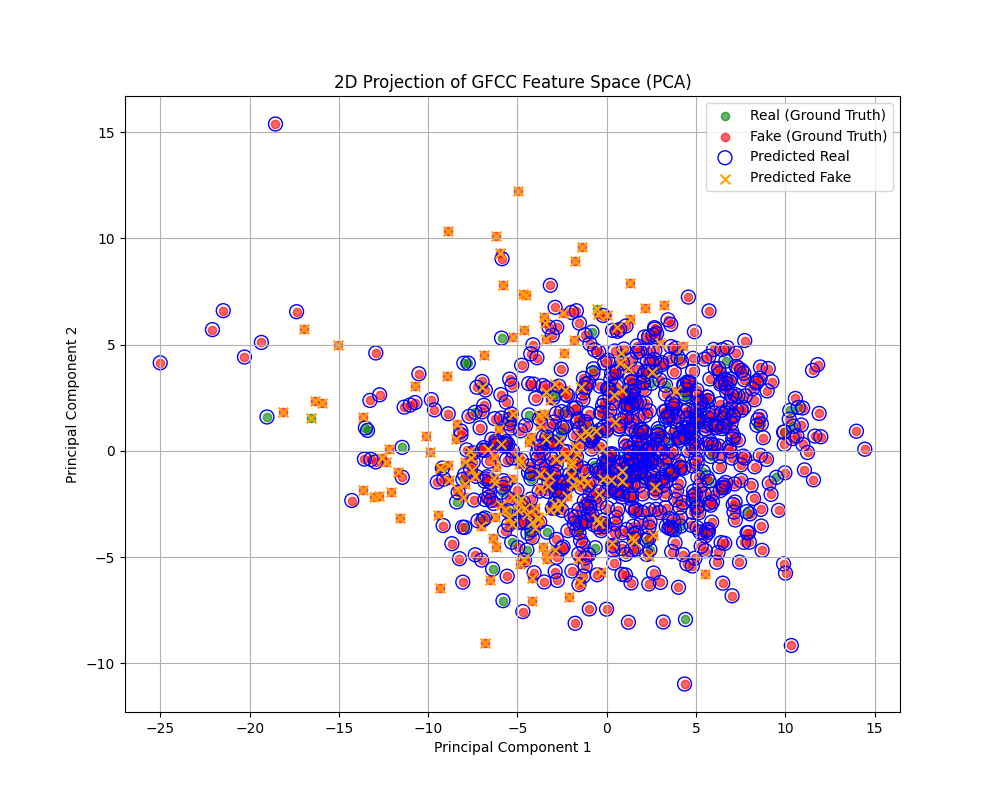}
\caption{Pop Goes the Verifier latent space}
\label{fs2_poison_synth}
\end{figure}
Figures \ref{clean_latent}, \ref{fs2_poison_vanilla}, and \ref{fs2_poison_synth} show the latent space embeddings after PCA. The clean model (Figure \ref{clean_latent}) maintains good separation between real and spoof classes. In contrast, the FS2 poison (Figure \ref{fs2_poison_vanilla}) causes severe model collapse: real and spoof embeddings heavily overlap with strong bias toward the “real” class and reduced variance along principal components. This collapse is even more pronounced with Pop Goes the Verifier (Figure \ref{fs2_poison_synth}), where embeddings form tighter clusters and occupy a narrower region.
The reduced variability stems from FS2’s inherent design for pronunciation stability, which limits temporal dynamics and frame-to-frame energy fluctuations critical for GFCC delta features. SyntheticPop injections further reinforce this consistency through deterministic pop envelopes and uniform noise, suppressing the natural spectral and amplitude variations that VoicePop relies on for discrimination.

\subsection{Bhattacharyya Results}
\label{stl:bht}

Unlike the waveform and spectrogram analysis of \ref{stl:ls}, which evaluates the visibility of injected perturbations in the signal itself, the Bhattacharyya score measures statistical similarity of an audio sample's bispectral characteristics to human and synthetic speech distributions. These analyses measure different properties of the attack and therefore are not expected to rank methods identically. 

The Bhattacharyya score quantifies how closely an audio sample’s bispectral features (magnitude and phase statistics) resemble human versus synthetic speech distributions. A pre-trained classifier produces a probability vector over human/synthetic classes. The final score is computed as 
\[ S_{\text{file}} = 1 - \frac{D_{\text{human}}}{D_{\text{human}} + D_{\text{synthetic}}} \] 
where \(D\) is the simplified Bhattacharyya distance to the empirical mean of each class. Higher values (closer to 1) indicate greater similarity to human speech.

\begin{table}[htbp]
\caption{Average Bhattacharyya Scores}
\label{tab:Bhattacharyya_scores}
\centering
\small
\begin{tabular}{|c|c|c|c|}
\hline
\textbf{Experiment} & \textbf{Human} & \textbf{Synthetic} & \textbf{Overall} \\
\hline
Human Audio (Real)        & 0.7988 & 0.6835 & 0.7411 \\
\hline
FastSpeech2 (FS2)         & 0.6448 & 0.4853 & 0.5651 \\
\hline
Pop Goes The Verifier     & 0.6476 & 0.4076 & 0.5276 \\
\hline
SyntheticPop & 0.7356 & 0.4403 & 0.5879 \\
\hline
\end{tabular}
\end{table}

\vspace{0.8em}

Both FS2 and Pop Goes the Verifier significantly lower the overall Bhattacharyya score compared to real human audio, indicating reduced similarity to human spectral patterns. Pop Goes the Verifier achieves the lowest Bhattacharyya score (0.5276), indicating that the added synthetic pops and Gaussian masking reduce bispectral similarity to human speech more than vanilla FS2. Therefore, although the method provides improved resistance to signal level inspection, it does not improve statistical similarity when viewed from this metric.

\section {Discussion}
This work highlights fundamental vulnerabilities in phoneme-based anti-spoofing systems such as VoicePop. Because these defenses rely heavily on detecting low-frequency pop bursts and spectral dynamics around phonemes, they are inherently susceptible to carefully crafted phoneme-mimicking attacks like SyntheticPop. A clear trade-off exists between poisoning effectiveness and resistance to signal level inspection. The initial SyntheticPop approach proved highly effective at poisoning the classifier but introduced more detectable artifacts. In contrast, the Pop Goes the Verifier variant trades some poison effectiveness for improved resistance to waveform and spectrogram inspection by embedding perturbations that better preserve audio naturalness while still degrading model performance. However, the Bhattacharyya analysis demonstrates that this does not correspond to greater statistical similarity to human speech.

Several defense strategies can mitigate such attacks, including adversarial training with poisoned samples, spectral anomaly detection for structured artifacts, improved data provenance checks, and the use of ensemble features beyond GFCC alone. However, these defenses were not evaluated in the present paper and therefore should be regarded as candidate mitigation strategies rather than validated countermeasures. Future work could empirically evaluate their effectiveness against both SyntheticPop and Pop Goes the Verifier. 

This study is also subject to several limitations. The evaluation was conducted using an even-training subset of the ASVSpoof dataset, a single ASV system (VoicePop), and a white-box supply chain threat model. The white-box setting was intentionally selected to isolate the feature-level vulnerability of phoneme-based liveliness detection under controlled conditions. Despite this, it does not evaluate attack transferability across different ASV architectures or scenarios where the adversary lacks knowledge of the target model. Additionally, the study does not include a formal human perception evaluation, statistical significance testing, hyperparameter sensitivity analysis, or robustness evaluation across varying poisoning budgets. Future work should therefore extend the evaluation to additional ASV systems, black-box threat models, lower poisoning budgets, and empirical assessments of both attack robustness and proposed defenses.

Although the assumed poison budget facilitates direct comparison with prior work, practical supply chain attacks may involve substantially smaller fractions of compromised training data. Examples include open source speech datasets, 3rd party dataset contributions, compromised preprocessing pipelines, or malicious data providers. Further work should therefore evaluate the effectiveness of proposed attacks under smaller poisoning budgets and more realistic data compilation scenarios. 

Collectively, these limitations should be interpreted as defining the scope of the present work rather than diminishing observed vulnerabilities. The objective of the study is to determine the feasibility of phoneme-aligned poisoning against phoneme-based liveliness detection under controlled conditions. Extending the evaluation to additional ASV architectures, alternative threat models, statistical robustness analyses, and empirical defense evaluation represents the natural next step. 

\section{Conclusion}


This research can be summarized as follows: a) The proposed Pop Goes the Verifier still allows for a respectable poison, as system accuracy drops 11.86\%. While not a significant or substantial poison like SyntheticPop, by being more strategic about the injected poisoning, the odds of being detected are much more compared to the total model collapse of SyntheticPop. b) In order to develop a data poison that can cripple the decision-making process in VoicePop, modifying FS2 with the injection of Pop Goes the Verifier leads to a moderately effective poison due to the added artifacts aiding with the low variability of FS2. As a result of the additional structured artifacts being attached to an already stable speech synthesis model, it caused even greater latent space collapse regarding decision-making. Additionally, the Gaussian noise overlay reduced the visual prominence of injected energy signatures during waveform and spectrogram inspection, providing an additional camouflage layer against signal-level inspection. This should not be interpreted as increased statistical similarity to human speech, as demonstrated by the Bhatachaya score. Lastly, the injected SyntheticPops disrupted the GFCC feature extraction process through the introduction of low-frequency perturbations, which interfered with energy deltas and distortion of changes in amplitude, thus interrupting the process for learning spoof patterns. Perhaps what is the biggest takeaway of this paper is that with all attacks benchmarked, it is clear that the sensitivity to pop noises that allow phoneme-based liveliness detection is also its primary vulnerability. Future work will evaluate the proposed strategies across additional ASV architectures and more realistic black-box threat models to determine whether the observed feature-level vulnerability generalizes beyond VoicePop. 

\section*{Acknowledgments}

Research reported in this publication was supported by the Division of Research and Innovation at San José State University under Award Number 25-RSG-08-134. The content is solely the responsibility of the author(s) and does not necessarily represent the official views of San José State University.

\bibliographystyle{IEEEtran}
\bibliography{references}

\end{document}